\documentclass[10pt, letterpaper]{iopart}

\usepackage[dvips]{graphicx, color}
\usepackage{latexsym} 
\usepackage{amsthm}
\usepackage{setstack}
\usepackage{amscd,amssymb} 
\usepackage{mathrsfs} 
\usepackage{cite}

\begin{document}

\title{Coupling Non-Gravitational Fields with Simplicial Spacetimes }
\author{Jonathan R. McDonald$^{1, 2}$ and Warner A. Miller$^{1}$}
\address{$^{1}$Department of Physics, Florida Atlantic University, Boca Raton, Florida 33431,  USA\\
$^{2}$  Institute for Applied Mathematics, Friedrich Schiller University, 07743 Jena, Germany }
\ead{jonathan.mcdonald@uni-jena.de}

\begin{abstract}

  The inclusion of source terms in discrete gravity is a long-standing
  problem.  Providing a consistent coupling of source to the lattice
  in Regge Calculus (RC) yields a robust unstructured spacetime mesh
  applicable to both numerical relativity and quantum gravity. RC
  provides a particularly insightful approach to this problem with its
  purely geometric representation of spacetime.  The simplicial
  building blocks of RC enable us to represent all matter and fields
  in a coordinate-free manner.  We provide an interpretation of RC as
  a discrete exterior calculus framework into which non-gravitational
  fields naturally couple with the simplicial lattice.  Using this
  approach we obtain a consistent mapping of the continuum action for
  non-gravitational fields to the Regge lattice.  In this paper we
  apply this framework to scalar, vector and tensor fields. In
  particular we reconstruct the lattice action for (1) the scalar
  field, (2) Maxwell field tensor and (3) Dirac particles.  The
  straightforward application of our discretization techniques to
  these three fields demonstrates a universal implementation of
  coupling source to the lattice in Regge calculus.
\end{abstract}
\pacs{04.60.Nc, 02.40.Sf}
\submitto{\CQG}

\maketitle

\newtheorem{assumption}{Assumption}
\newtheorem{define}{Definition}
\newtheorem{theorem}{Theorem}
\newtheorem{prop}{Property}

\section{Non-gravitational Source in Simplicial Spacetime}
\label{sec:1}

Regge Calculus (RC) is a discrete coordinate-free formulation of
Einstein's geometric theory of gravitation \cite{Regge:1961}.  Its
simplicial spacetime directly incorporates the local Poincar\'{e}
invariance of General Relativity into finite domains \cite{CDM:1989}.
The simplicial geometry is by now well understood \cite{Regge:2000},
but how is one to couple non-gravitational sources to this lattice?
We are unaware of a completely unified approach to the incorporation
of matter into Regge calculus (RC) which utilizes the natural
geometric structure and locally finite implementation of both the
Poincar\'{e} and diffeomorphism symmetry.  In this manuscript we
provide such a description.

While the inherent representation of a local tangent space within each
simplex provides us with an {\em a priori} orthogonal and holonomic
frame, i.e. locally flat Minkowski metric structure with finite
domain, the way in which we couple source to this structure must also
utilize the discrete diffeomorphism invariance \cite{Miller:1986,
  McDonald:2008c}. It is this invariance that automatically enforces
the conservation of stress-energy at each vertex in the lattice.  This
is equivalent to identifying the local structure on which fields must
be embedded so as to be consistent with the automatic preservation of
conservation of source \cite{MTW:1973}.  In this manuscript we provide
a unified approach to constructing non-gravitational fields in RC
based on the interpretation of the Regge lattices as discrete
differential forms \cite{Whitney:GeomInt, Desbrun:2005DEC}.  We begin
in Sec.~\ref{sec:2} by outlining the basics of a discrete exterior
calculus.  We then apply this framework to the Klein-Gordon
(Sec.~\ref{sec:kg}), Maxwell' (Se.~\ref{sec:em}), and the Dirac fields
(Sec.~\ref{sec:df}).

\section{Discrete Differential Forms}
\label{sec:2}

In order to accurately retain the geometric properties of fields in a
discrete framework, one must embed, in a consistent way, the fields
into the lattice through a correspondence between the geometry of the
lattice and field variables \cite{Whitney:GeomInt, Arnold:2006}.  This
has been the primary focus of methods in computational
electromagnetism \cite{Bossavit:1991, Arnold:2006, Stern:2008} where
the focus is on demonstrating increased stability and automatic
retention of internal symmetries.  In these approaches to the
discretization of Maxwell's equations the electromagnetic
four-potential, $A^{\mu}$ is constructed from chains of 1-forms that
naturally emerge in the underlying triangulation \cite{Stern:2008}.
The intuitive notion is that one projects a continuous $k$-form onto
an algebraic $k$-chain determined by the $k$-simplexes of the lattice.
In this manuscript we provide a cursory overview of this construction
and its geometry.  A more complete discussion of the implementation of
discrete exterior calculus is available in Desbrun {\em et al.}
\cite{Desbrun:2005DEC} or Arnold {\em et al.}  \cite{Arnold:2006}.

The piecewise-flat simplicial lattice of RC is an image of a
curvilinear simplicial complex on a continuous manifold under an
appropriate isomorphism.  We use this map to project the $k$-forms
onto the simplicial complex and identify its projections into the
lattice.  We start be explicitly constructing such simplicial forms.
If $\omega$ is a $k$-form then we define the lattice representation of
$\omega$ by,
\begin{equation}
\label{fmap}
\Pi(\omega, {\sigma}^{k}):=\left< \omega | {\sigma}^{k}\right>  = \int_{\sigma^{k}} \omega,
\end{equation}
where $\sigma^k$ is a $k$-dimensional simplex. This ``projection
inner-product'' maps each continuum form into its corresponding
piecewise-flat simplicial form.  The key feature of the discrete
exterior calculus approach that is captured by Eq.~\ref{fmap} is that
any physical field contributes scalar weights to the simplicial
differential forms naturally supplied by the lattice.  This greatly
simplifies any calculations on the simplicial net as one need only
work with $k$-simplexes and their associated scalar value.  This
closely mimics the fundamental approach of RC to gravitation in which
geometric observables are defined via weighting of appropriate
$k$-simplexes with scalar weights \cite{CFL:1982b, Miller:1997}.

In addition to the discrete differential forms one must introduce a
Hodge operator to map simplicial forms to dual forms in the
lattice. The construction of such an operator is not unique
\cite{Stern:2008} as one can arbitarily construct a consist mapping
based on circumcenters, incenters, barycenters, etc. Despite the
ambiguity in constructing a dual lattice, there are distinct
advantages to utilizing the circumcentric dual.  It is this dual that
appears to be most natural in RC \cite{CFL:1982a, CFL:1982b,
  FFLR:1984, FL:1984 , Regge:2000, Miller:1986, McDonald:2008c}
because of its orthogonality to the simplicial lattice. This
orthogonality provides a factorization of $d$-volumes associated with
any given $k$-simplex into orthogonal components which lie in either
the simplicial or dual lattice.  This factorization of the homology
(Delaunay lattice, $\sigma$) and the co-homology (Voronoi lattice,
$*\sigma$) was essential in defining the Einstein-Hilbert action in RC
\cite{Miller:1997} and the vertex-based scalar curvature
\cite{McDonald:2008b}.  However, one may choose to define the dual in
any consistent manner.

Moreover, to ensure that the dual is a well-defined
\cite{McDonald:Voronoi} lattice we restrict ourselves to Delaunay
simplicial lattices and their circumcentric Voronoi duals.  This
restriction is a natural assumption since we wish to require that any
local topology constructed from the hybrid $d$-volumes remains
Hausdorff.  The Hodge dual in a simplicial $d$-dimensional geometry
thus maps $k$-forms from $k$-simplexes to $(d$-$k)$-forms on the
$(d$-$k)$-faces of the circumcentric Voronoi dual.  Explicitly this
can be defined as \cite{Stern:2008},
\begin{equation}
\label{eq:Hodge}
\frac{ 1}{|\sigma^{k}|}\left< \alpha| \sigma^{k}\right>= {\rm Sign}(\sigma^{k})\frac{1}{|{*}_{\sigma^{k}}|} \left< *\alpha| *\sigma^{k}\right>,
\end{equation}
where $|\sigma^{k}|$ is the volume of k-simplex, $\sigma^{k}$, and
where one includes an appropriate sign based on whether the
$k$-simplex is space-like or time-like.  This allows us to explicitly
define the right-hand-side of Eq.~\ref{eq:Hodge} as a mapping from
$*\alpha$ to elements, $*\sigma^{k}$, of the dual lattice
\begin{equation}
\Pi(*\alpha):= \left<*\alpha|*{\sigma}^{k}\right> =\int_{*\sigma^{k}} *\alpha.
\end{equation}
The assignment of $p$-forms $\omega \in \Lambda^{k}$ to either the
simplicial or the dual lattice is somewhat arbitrary.  For the sake of
clarity we examine the geometric meaning of those elements of
$\Lambda^{k}$ to identify such an assignment.  In this spirit, a
$1$-form is equivalent to a covector, and a general $k$-form is given
as the totally anti-symmetric product of $k$ 1-forms.  The
identification of $1$-forms with covectors indicates that the dual of
a $(d$-$1)$-form is identified as a vector on a manifold.  We thus make
the identification that forms on the circumcentric dual represent the
standard $k$-forms while the elements of the simplicial lattice
represent the dual space.  For further clarity, we identify simplicial
forms as $k$-forms in $^{*}\Lambda^{d-k}$ and dual forms, i.e.  forms
represented on the dual lattice, as $k$-forms in $\Lambda^{k}$.

With the definition of the dual forms we can introduce an inner-product of two discrete forms on the manifold. 
\begin{eqnarray}\label{eq:funcprod}
(\omega, \eta) &:= \int \left< \omega | \eta\right> = \int \omega \wedge *\eta   &    \;\;\;\;\;\;\;\;\;\;\; ({\rm continuum}) \nonumber \\
   (\omega, \eta) &:= \sum_{\sigma^{k}} \left[\omega \wedge * \eta\right]_{\sigma^{k}}  = \sum_{\sigma^{k}} \left< \omega| \eta\right> V^{(d)}_{\sigma^{k}} = \sum_{\sigma^{k}} \omega (\sigma^{k}) \eta (\sigma^{k}) V_{\sigma^{k}}^{(d)}  &  \;\;\;\;\;\;\;\;\;\;\; ({\rm lattice}). 
\end{eqnarray}
Here, $V^{d}(\sigma^{k})$ is the hybrid between the simplicial
$k$-form and its dual.  The circumcentric dual gives a particularly
simple form for these volumes,
\begin{equation}
V_{\sigma^{k}}^{(d)} =\frac{1}{{d \choose k}}|\sigma^{k}| \cdot |*\sigma^{k}|,
\end{equation}
resulting from the inherent orthogonality between the two lattices. We
should also distinguish between the functional inner-product, $(
\cdot, \cdot)$, and the spacetime inner-product, $\langle \cdot |
\cdot\rangle$, in that the latter acts locally on the simplicial
skeleton of the manifold while the former tells us how to ``integrate"
discrete forms over the entire manifold.

The discrete exterior derivative maps a $k$-simplicial (dual) form
mapped on $\sigma^{k}$ ($*\sigma^{d-k}$) into a $(k$+$1)$-simplicial
(dual) form.  This implies a discrete sum over the $k$+$1$ simplexes
(dual cells) incident on $\sigma^{k}$ ($*\sigma^{d-k}$).  To clarify
this we rely on the exterior co-derivative which maps a $k$-simplicial
(dual) form into a $(k$-$1)$-simplicial (dual) form.  The exterior
derivative and co-derivative are related through the inner-product as
usual
\begin{eqnarray}
\left<{\bf d}\alpha   | {\sigma}^{k}  \right> = \frac{1}{|\sigma^{k}|} \left< \alpha  | \delta \sigma^{k}  \right>   & \;\;\;\;\;\&\;\;\;\;\;&  \left<{\bf d}\alpha   | *{\sigma}^{k}  \right> =  \frac{1}{|*\sigma^{k}|} \left< \alpha  | \delta (*\sigma^{k})\right>.
\end{eqnarray}
where $\delta := *{\bf d}*$.  We thus have a discrete representation
of Stoke's Theorem, which leads us to the discrete (co-)boundary of a
(co-) boundary principle
 \begin{eqnarray}
{\bf d}{\bf d} \alpha \equiv 0   & \;\;\;\;\;\;\;\; \&  \;\;\;\;\;\;\;\;& {\bf \delta}{\bf \delta} \alpha \equiv 0.
\end{eqnarray}    
This provides the necessary de Rham cohomology with the spaces
${}^{*}\Lambda^{k}$ and $\Lambda^{k}$ defined on the simplicial
lattice and its dual, respectively.

The discrete representation of differential exterior calculus
presented here, is by virtue of the projection operator, structurally
equivalent to continuum exterior calculus though defined on a
piecewise-flat manifold.  Using insight from the continuum
Kirchhoff-like conservation of stress-energy \cite{Wheeler:1972,
  Cartan:1928} we can explicitely show how this framework is already
expressed in RC \cite{McDonald:2008b}. This provides a robust
framework for coupling non-gravitational fields  to simplicial
spacetimes.  In the next three sections we provide a construction of
the scalar field action for scalar, vector and spinor fields.  In
particular we construct the electromagnetic field and the Dirac field
action for simplicial spacetimes with four dimensions.

%%%%%Scalar Fields
\section{Discrete Scalar Fields}
\label{sec:kg}

As an illustration of this construction we derive the lattice action
for a scalar field and show its equivalence to its traditional finite
difference representation \cite{Hamber:1994}.  Here we provide a
direct calculation of the discrete action from the continuum action
through a implementation of the discrete exterior forms of
Sec.~\ref{sec:2}.  The scalar field action in the continuum is given
by
\begin{equation}
S[\phi, \bar{\phi}] = \int d^{4}x \frac{1}{2}\left[  \partial^{\mu} \phi \partial_{\mu}\bar{\phi}  - m^{2} \bar{\phi}\phi\right] = \frac{1}{2}\left( {\bf d}\phi ,  {\bf d} \bar{\phi}\right) - \frac{m^{2}}{2} \left( \phi , \bar{\phi} \right).
\end{equation}
The differential form expression for the scalar field action contains
both 1-forms and 0-forms which we must embed in the lattice.  We first
identify whether these are simplicial forms or dual forms.  The
stress-energy corresponding to the propagation of the field is
directed along an edge of the lattice \cite{McDonald:2008c}.  To
maintain an appropriate identification of stress-energy with the field
we express flux of scalar field as directed along edges of the
simplicial lattice.  In other words, we express each 1-form in the
kinetic term of the action as projected onto the edges of the lattice.
Likewise, each 0-form in the mass term is projected onto the vertices
of the simplicial spacetime;
\begin{equation}
\Pi(\phi) = \left< \phi | v\right> = \phi(v),   
\end{equation}
\begin{equation}
\Pi({\bf d}\phi)=\left< {\bf d} \phi | {L} \right> = \sum_{v \subset L} \frac{1}{|L|} \left< \phi | v\right> = \frac{\phi(v+{\bf L})- \phi_{v}}{|L|}.
\end{equation}
This yields the standard finite differencing term for a scalar field on the lattice.   Using this coefficient for the 1-form directed along $L$,  we obtain the discrete action.  
\begin{eqnarray}
S[\phi, \bar{\phi}] &= \frac{1}{2}\left[ \sum_{L} \left< {\bf d}\phi | {\bf d}\bar{\phi}\right> V^{(4)}_{L} - m^{2} \sum_{v}\phi\bar{\phi} V^{(4)}_{v}         \right] \nonumber \\
& = \sum_{L}   \frac{1}{2} \frac{\phi(v+{\bf L})- \phi_{v}}{|L|} \cdot \frac{ \bar{\phi}(v+{\bf L})- \bar{\phi}_{v}}{|L|} \underbrace{\frac{1}{4}|L| V^{*}_{L}}_{V^{(4)}_{L}} - \frac{m^{2}}{2} \sum_{v}\phi(v)\bar{\phi}(v) {V^{*}_{v}}         \
\end{eqnarray}
Here we assume that $v$ is the base vertex for the oriented edge ${\bf
  L}$.  This coincides with the action suggested in \cite{Hamber:1994}
with the four-volumes given by the edge-based simplicial-circumcentric
dual hybrid volumes.  The stress-energy tensor of the scalar field is
a doubly-projected tensor along the edge $L$ with conservation given
by a Kirchhoff-like conservation law \cite{McDonald:2008c} .  This can
be explicitly and simply calculated using the orthogonal decomposition
of the hybrid cells.  The stress-energy tensor associated with a given
edge for the complex scalar field is
\begin{equation}
T_{LL}V^{*}_{L} := \frac{\delta S}{\delta L}[\phi, \bar{\phi}] = -\frac{1}{8} \left[  \frac{(\phi(v+{\bf L})- \phi_{v})(\bar{\phi}(v+{\bf L})+ \bar{\phi}_{v})}  {L^{2}} - \sum_{v \subset L} \frac{m^{2}}{2}\phi(v)\bar{\phi}(v)  \right] V_{L}^{*}
\end{equation}
with conservation of source given by
\begin{eqnarray}
\label{scl}
\sum_{L\supset v} T_{LL}  &= \frac{1}{8} \sum_{L\supset v}\left[  -\frac{(\phi(v+{\bf L})- \phi_{v})(\bar{\phi}(v+{\bf L})- \bar{\phi}_{v})}  {L^{2}} - \sum_{v' \subset L} \frac{m^{2}}{2}\phi(v')\bar{\phi}(v')  \right] \nonumber \\
\nonumber \\
&\approx 0
\end{eqnarray}
This provides us with a direct and minimal coupling of the complex
scalar field with the Regge lattice.  It also provides us with a
Kirchhoff-like expression for the conservation of scalar field
stress-energy.  This scalar field application provides a literal
interpretation of the Kirchhoff-like conservation law at each vertex,
$v$, as a flow of field along the edges of the simplicial lattice
meeting at $v$.  One can see explicitly in Eq.~\ref{scl} the change in
the field along each edge from one vertex to the another.  In the next
section (Sec.~\ref{sec:em}) we extend the simplicial exterior calculus
approach to Maxwell's equations where the field is vectorial, and the
correspondence with a vertex-based conservation law is not as
transparent as the scalar field example done here, instead we must
rely wholly on the framework set up in Sec.~\ref{sec:2}.
%%%%%  EM Field
\section{Simplicial Electromagnetic Fields}
\label{sec:em}

The electromagnetic field is defined by the connection 1-form,
$A^\mu$, and the action is given by the square of the curvature
2-form, $F^{\mu\nu}$,
\begin{equation}
S[A] = -\frac{1}{2} (F, F) = -\frac{1}{2} \int d^{4}x\;\; F \wedge *F = -\frac{1}{2} \int d^{4}x\;\; {\bf d}A \wedge *{\bf d} A.
\end{equation}  
We now restrict ourselves for the rest of the manuscript to four
spacetime dimensions in which we must represent the 2-form ${\bf d}A$
on a 2-simplex.  The connection 1-form $A$ is defined on the edges of
the simplicial lattice.  The curvature 2-form is therefore determined
entirely by the values of $A$ on its boundary;
\begin{equation}
\left< {\bf d} A| \sigma^{2} \right>  = \frac{1}{ |\sigma^{2}|} \langle A|\delta \sigma^{2}\rangle= \sum_{L\subset \sigma^{2}} \langle A| L\rangle = \frac{|L|}{ |\sigma^{2}|} \sum_{L \subset \sigma^{2}} A_{L}.
\end{equation}
Here we utilize the fact that we integrate over the entire boundary
which introduces a factor of $|L|$ into the contribution from each
edge.

It is assumed that the discrete forms are valued on the appropriate
$k$-simplex such that we need only work with the scalar values of the
coefficients.  Making the substitutions into the source-free Maxwell
Lagrangian, the simplicial electromagnetic action becomes
\begin{equation}
S[A] = -\frac{1}{4} \sum_{\sigma^{2}}  \left(  \sum_{L\subset \sigma^{2}}  \frac{|L|}{|\sigma^{2}|} A_{L}\right)^{2} \underbrace{ \frac{2}{d(d-1)} |\sigma^{2}| |*\sigma^{2}|}_{V^{(d)}_{\sigma^{2}}}
\end{equation}

Introducing source into the action requires the 1-form current ${\it
  j}$, whose dual is the 3-form current current density, $*{\it j}$,
defined on the 3-volumes dual to an edge.  The former determines the
direction of propagation of the current while the latter determines
the volume pierced by the flow of the current.  This introduces the
interaction in the action
\begin{equation}
 (A, {\it j}) = \sum_{L} \left\langle{\it j}_{L},A_{L}\right\rangle V^{(d)}_{L}
\end{equation} 
While, gauge invariance of the potential 1-form under the
transformation $A + {\bf d}\Lambda$ is automatically satisfied in the
source-free terms of the action, invariance of the action should give
indications of the conservation of source based on the transformations
of the source terms.  Under the gauge transformation, the source terms
of the action give
\begin{eqnarray}
\label{emaction}
\left( A+ {\bf d}\Lambda, {\it j}\right)& = \left( A, {\it j}\right) + \left( {\bf d} \Lambda, {\it j}\right)  \nonumber \\
&=  \left( A, {\it j}\right) +\sum_{L} (\Lambda_{v+L} - \Lambda_{v})\cdot {\it j_{L}} \nonumber \\
&=  \left( A, {\it j}\right) -\sum_{v} (\Lambda_{v})\sum_{L\supset v} {\it j_{L}}.
\end{eqnarray}
Under arbitrary gauge transformations, the invariance of the action implies
\begin{equation}\sum_{L \supset v} {\it j}_{L} = \sum_{*L \subset \partial (*v)} {\it *j}_{L}= 0.
\end{equation}
where the first sum gives the Kirchhoff-like conservation of the
current 1-form, while the second sum is based on the more typical
representation of the divergence of the current density 3-form.  This
result was first shown in \cite{Sorkin:1975} where the projection of
the four-potentional is equivalent to the construction here.

This construction follows a path to that of the Sorkin construction
\cite{Sorkin:1975}.  However, by using only discrete differential
forms we express the electromagnetic field in coordinate-free language
determined by scalar weightings of simplicial elements.  We thus avoid
defining tangent-space-valued tensors in any strict sense as appears
to be the base nature of gravitational variables in RC.

While the action (Eq.~\ref{emaction}) is similar in character, it is
explicitly distinct from the Sorkin approach to electromagnetism on
the spacetime mesh \cite{Sorkin:1975}.  Sorkin's approach to the
problem utilizes the full tangent space associated a $d$-simplex to
determine the affine components of the Maxwell tensor locally.  The
method presented here extends Sorkin's original embedding of $A^{\mu}$
into the lattice to incorporate the more recent understandings of the
dual lattice structure inherent in RC \cite{CFL:1982a, CFL:1982b,
  FFLR:1984, FL:1984, Miller:1986, Miller: 1997, McDonald:2008b,
  McDonald:2008c}.  Using the dual lattice, one already finds that the
spacetime curvature automatically decomposes into hinge-based
curvature tensors proportional to the scalar curvature.  In
constructing the action, these hinge-curvatures are weighted by the
hybrid volumes described in Sec.~\ref{sec:2} \cite{Miller:1997} as the
determiners of the local measure on the lattice.  This is a result of
the projection of the continuum curvature 2-form onto the
2-dimensional polygonal faces of the dual lattice, the carriers of
information of spacetime curvature.  Moreover we have shown
\cite{McDonald:2008b} how this construction leads to alternative
descriptions of curvature at localized sites in the lattice.  The
simplicial electromagnetic action outlined above follows in the same
spirit as curvature's manifestation in RC.  The Maxwell tensor is
locally distributed on the 2-forms of the simplicial net and weighted
by the corresponding hybrid volumes.  This allows us the flexibility
of stringing together chains of simplicial 2-forms to define locally
simple Maxwell curvature in the simplicial lattice, as was shown
explicitly for spacetime curvature in RC \cite{McDonald:2008b}.

%%%% It is useful to note that spirit of the action
%%%% (Eq.~\ref{emaction}) and definition of field variables agrees
%%%% with the earlier treatment by Sorkin\cite{Sorkin:1975}.  However,
%%%% this identification of field strength with 2-simplexes and the
%%%% definition of the action as a sum over 2-simplexes differs from
%%%% the Sorkin simplex-based action.  There it was necessary and
%%%% appropriate to directly sum over quantities in a given simplex as
%%%% the affine coordinates were only valid in a given simplex, and
%%%% thus the weighting assigned to the trace of the curvature 2-form
%%%% is given by the volume of the simplex.  In contrast, this
%%%% formulation corresponds directly with modern techniques in
%%%% computational electromagnetism \cite{Bossavit:1991,
%%%%   Stern:2008}. When we apply this method to a flat spacetime
%%%% lattice we recover the the computational electromagnetism
%%%% formulation exactly.  %%%%%%%%%%

The discrete differential form approach to defining a field on a
simplicial lattice for the two cases we have examined illustrate the
construction and its usefulness for non-gravitational fields in RC.
We have been able to construct analogous actions directly from the
continuum action via projection of the continuum fields onto the
corresponding elements of the lattice.  The scalar field, in
particular, provides an identical match to previous literature on
scalar fields in RC.  The action for the electromagnetic field differs
from previous descriptions as we no longer require an explicit
tangent-space-valued tensor to be constructed interior to each
simplex.  This compares well with computational electrodynamic
applications for flat spacetimes.  In the next section we extend this
formalism and derive an action for fermionic fields that is consistent
with the fundamental coordinate-free structure of and the flow of
stress-energy in RC.

\section{The Simplicial Dirac Field}
\label{sec:df}
%%%%%Spinor Field (using the clifford algebra of differential forms)

Before we derive our version of the Dirac action in RC it is important
to note some of the key elements that go into such a construction.
These elements are pulled together from considerations of (1) the
Dirac field itself, and from (2) the simplicial lattice geometry.  The
first component is the need for an orthogonal frame for the
construction of a Dirac spinor.  This is fundamentally related to the
need to provide local representations of the Dirac
$\gamma$-matrices. These matrices form their own basis in the Clifford
algebra of spacetime.  Therefore, one must be able to construct an
isomorphism between the representations of the geometry of the
spacetime (the basis 1-forms) and the $\gamma$-matrices
\cite{GS:1987}.  In RC this is quite straightforward since we are
provided with finite representations of spacetime whose basis 1-forms
can be used to construct an orthonormal and holonomic frame.  However,
this cannot be taken too literally. As with the other fields, the
Dirac field is a spacetime-valued field.  Moreover, in light of the
conservation of stress-energy as derived in \cite{McDonald:2008c}, it
is known that the flow of stress-energy is directed on the edges of
the simplicial lattice.  This should give some indications that the
flow of field (or particles), as the carrier of stress-energy, is also
directed along the edges of the simplicial lattice.  In the
construction given below we follow this path by defining the Dirac
field on the simplicial geometry.

The standard construction of fermionic fields uses an action given by
\begin{equation}
S[\psi, \bar{\psi}]= \int d^{4} x \; \; i\bar{\psi}\gamma^{a}e^{\mu}_{a}\nabla_{\mu}\psi  - m \bar{\psi}\psi,
\end{equation}
where $e^{\mu}_{a}=(e^{a}_{\mu})^{-1}$ is the co-tetrad connecting a
holonomic frame defining the covariant derivative to the orthonormal
frame representation of $\gamma^{a}$.  Before we assign a simplicial
form to the fields directly we note that the field $\psi$ is
explicitly based in a representation of the double-cover ${\rm SL}(2,
\mathbb{C})$ and not in a representation of ${\rm SO}(3, 1)$.  In
order to make the connection with our framework more clear we write
the Dirac action in a more suggestive form,
\begin{equation}
S[\psi, \bar{\psi}]=  i\bar{\psi}(\gamma^{a}e_{a}, D )\psi  - m (\bar{\psi},\psi). 
\end{equation}
where $D$ is the covariant exterior differential.  Here we view the
first term as an inner-product between operators for purely
illustrative purposes.  One could also expand this first term into two
by including the fields $\psi$ and $\bar{\psi}$ in the inner-product
with $D$ acting on $\psi$ in one term and $D$ acting of $\bar{\psi}$
in the second.  To keep the notation compact we will assume the first
term in the action to imply
\begin{equation}
i\bar{\psi}(\gamma^{a}e_{a}, D )\psi  =  i(\bar{\psi}\gamma^{a}e_{a}, D\psi ) - i({\psi}\gamma^{a}e_{a}, D\bar{\psi} ).
\end{equation}
In this representation of the action it is evident how one must embed
the field into the lattice.  The co-tetrad is naturally the simplicial
1-form since the tetrad is a dual 1-form (defining the tangent space
interior to a simplex).  Moreover, the covariant exterior derivative
maps a 0-form to a 1-form on the simplicial lattice. The 0-form nature
of the field suggests that the each vertex, $v$, of the simplicial
lattice is the natural placeholder for the fields., i.e. without
specifically assigning a basis or tangent space in which to represent
the field.  Rather we are free to choose any simplex meeting at $v$ to
determine the tangent space for the field.  Once a tangent space is
chosen, the {\em vielbein} explicitly determines the spinor basis and
matrix representation of the Clifford algebra of the $\gamma^{a}$ in
the standard way \cite{GS:1987, Penrose:spinors}.  The term
$\gamma^{a}e_{a}$ is then given by the projection $|L| \cdot
\Pi(\gamma) = \gamma_{L}$ which assigns a Dirac matrix representation,
in some appropriate basis, to the edges of the lattice. Since each
term in the action is a Lorentz scalar, we can make the choice of
basis independently for each term in the discretized action.
\begin{equation}
S[\psi, \bar{\psi}]=\sum_{L}  \left(i\left<\bar{\psi}_{L} \gamma_{L} |D_{L}\psi\right> + c.c. \right)V_{L}^{*} + \sum_{v} m\bar{\psi}(v)\psi(v) V^{*}_{v}.
\end{equation}
Here we take the value of the field on a given edge, $\psi_{L}$, to be
the average field on that edge.
\begin{equation}
\psi_{L} = \frac{\psi(v+L) + \psi(v)}{2}.
\end{equation}
The ability to freely choose an appropriate tangent space greatly
simplifies the expression of the action.  In particular, since each
edge has multiple simplexes hinging on the edge, we can define any
edge-based term of the action in the tangent space of one such
simplex.  This reduces the covariant derivative of the field to a
flat-space differential.  As with the scalar field the derivative
terms reduce to finite-differencing terms and we obtain a simplified
discrete action without the need for spin connections.
\begin{equation}
S[\psi, \bar{\psi}]=\sum_{L}  \left(i\bar{\psi}_{L} \gamma_{L} \frac{\psi(v+L) - \psi(v)}{L} + c.c. \right)V_{L}^{*} + \sum_{v} m\bar{\psi}(v)\psi(v) V^{*}_{v}.
\end{equation}
This local action utilizes the the freedom to assign each flat tangent
spaces to its fullest degree.  This in turn reduces all derivatives
and finite differences to their flat-space representations.  This
should simplify local calculations of the fields on the lattice.
However, for non-local calculations using the Dirac field, one will
necessarily introduce a spin-connection to transform one
representation of the Clifford algebra to another.  These are given
explicitly by orthogonal transformations of the vielbein from one
simplex to another.  Geometrically this is viewed as transport along
the edge dual to the boundary between two neighboring simplexes. A
short discussion of this transformation is given in \cite{Ren:1988}
where the Dirac field is represented as a the dual-vertex-based
action.

%%%Discussion
\section{A Unified Approach to Conservation of Stress-Energy}

The inclusion of sources into RC has been an area of active research
for some years with various approaches modeled without a unified
principle determining the implementation.  This manuscript has
provided an integrated framework based on differential forms for the
embedding of fields into a simplicial lattice.  The developments in
computational electromagnetism using an approach called discrete
exterior calculus (or discrete differential forms) together with our
earlier results have led us to a formal theory for projecting smooth
differential forms into a discrete pseudo-Riemannian lattice.  The
utility of this approach is its dependence only on the lattice
structure and geometry, and its explicit lack of dependence on
coordinate systems.

For scalar and fermion fields, the actions retain similarity to the
more familiar finite-difference schemes on the simplicial lattice with
appropriate volume weighting. For the Dirac field, we took a
minimalistic approach based on the conventional quantum field theory
construction of Dirac spinors.  It might prove useful to also
investigate a K\"{a}hler-Dirac formalism for spin-$\frac{1}{2}$
particles which explicitly uses the Clifford algebra of differential
forms \cite{GS:1987}.  We leave this to future work.  The construction
of the electromagnetic field outlined in this manuscript follows the
spirit of RC by embedding the fields without explicit reference to
affine coordinate systems inside simplexes.  The fields are directly
encoded only on the edges and triangles of the lattice with dependence
given by the incidence matrix of the lattice.  By encoding the fields
onto the simplicial skeleton we have also ensured a direct connection
with the conservation of stress-energy on the lattice
\cite{McDonald:2008c}.  The Kirchhoff-like conservation principle
tells us that the flow of stress-energy is directed along the edges of
the lattice.  This topological formulation supports the assumption
that each field should be encoded onto the skeleton as we have
suggesteed.  Encoding source fields on the simplicial skeleton
provides a concrete and geometric connection between the field and its
conserved current, i.e. stress-energy, under diffeomorphisms.

In canonical RC there have been a variety of approaches to embedding
non-gravitational fields in the simplicial lattice.  To date, these
approaches have been based on finite-difference schemes directly
applied to the lattice or barycentric coordinate representations of
field tensors.  Fields have been encoded in either the simplicial
skeleton or dual skeleton depending on convenience of representation.
However, a universal approach towards embedding non-gravitational
sources into RC with due regard to conservation of stress-energy has
so far been absent from the literature.  The work presented here
identifies a framework for coupling source to field, in accordance
with automatic conservation of source, in canonical RC.  Moreover, the
representations of fields and their actions are do not require
coordinization in the simplicial blocks.  Instead we assign
appropriate scalar or spin-valued weightings to elements of the
simplicial skeleton through projections from the continuum
representation.

In encoding the fields in the simplicial geometry based on canonical
RC, we define a universal prescription for assignment of field to the
lattice.  This provides a foundation for carrying over these
derivations to model-dependent descriptions of simplicial spacetimes.
Reliance on only the canonical structure of RC without regard to
specific dynamical behavior allows this approach to be universally
applicable to numerical relativity or simplicial models of quantum
gravity.  However, appropriate modifications to the dynamics of the
source fields may well be necessary to ensure consistency with
application to a given model of quantum gravity.  Such analysis of
these foundations in the context of specific models of quantum gravity
is a topic for future investigation.

\section*{Acknowledgements} 

We would like to thank Chris Beetle, Seth Lloyd and Ruth Williams for
some fruitful discussions on these matters.  We also thank Stephon
Alexander for reinvigorating our interest in fermion fields in RC.
One of us (JRM) would also like to acknowledge partial support from
NSF grant 0638662 and DFG grant SFB/TR7  ``Gravitational Wave
Astronomy''.

\end{document}